\documentclass[12pt]{article}
\pdfoutput=1

\usepackage{fixltx2e,fix-cm}
\usepackage{amssymb}
\usepackage{amsmath}
\usepackage{graphicx}
\usepackage{makeidx}
\usepackage{multicol}
\usepackage{a4wide}
\usepackage{amsthm, amsmath, amssymb}
\usepackage[square,numbers]{natbib}
\usepackage[singlelinecheck=off]{caption}

\newtheorem{example}{Example}

\usepackage{tikz}
\usepackage{subfigure}
\usepackage{graphicx}

\usetikzlibrary{shapes,arrows,fit}

\usepackage{verbatim}

\def\ci{\perp\hspace{-6pt}\perp}
\def\nci{\perp\hspace{-10pt}\slash\hspace{-9.5pt}\perp}

\title{A review of some recent advances in causal inference}
\author{Marloes H. Maathuis and Preetam Nandy}
\date{}

\begin{document}

\maketitle



\tableofcontents

\section{Introduction}\label{intro}

Causal questions are fundamental in all parts of science. Answering such questions from observational data is notoriously difficult, but there has been a lot of recent interest and progress in this field. This paper gives a selective review of some of these results, intended for researchers who are not familiar with graphical models and causality, and with a focus on methods that are applicable to large data sets.

In order to clarify the problem formulation, we first discuss the difference between causal and non-causal questions, and between observational and experimental data. We then formulate the problem setting and give an overview of the rest of this paper.

\subsection{Causal versus non-causal research questions}\label{sec: causal vs non-causal questions}

We use a small hypothetical example to illustrate the concepts.

\begin{example}\label{ex: prisoners}
   Suppose that there is a new rehabilitation program for prisoners, aimed at lowering the recidivism rate.
   Among a random sample of 1500 prisoners, 500 participated in the program. All prisoners were followed for a period of two years after release from prison, and it was recorded whether or not they were rearrested within this period. Table  \ref{table: prisoners} shows the (hypothetical) data.  We note that the rearrest rate among the participants of the program ($20\%$) is significantly lower than the rearrest rate among the non-participants ($50\%$).

   \begin{table}[h]
      \caption{Hypothetical data about a rehabilitation program for prisoners.}%
      \begin{tabular}{lccc} \hline
          &\textbf{Rearrested} &\textbf{Not rearrested} &\textbf{Rearrest rate}\\ 
      Participants & 100 & 400   & 20\% \\
      Non-participants & 500 & 500 & 50\%\\\hline
      \end{tabular}
      \label{table: prisoners}
   \end{table}
\end{example}

We can ask various questions based on these data. For example:
\begin{enumerate}
  \item Can we predict whether a prisoner will be rearrested, based on participation in the program (and possibly other variables)?
  \item Does the program lower the rearrest rate?
  \item What would the rearrest rate be if the program were compulsory for all prisoners?
\end{enumerate}

Question 1 is \emph{non-causal}, since it involves  a ``standard" prediction or classification problem. We note that this question can be very relevant in practice, for example in parole considerations. However, since we are interested in causality here, we will not consider questions of this type.

Questions 2 and 3 are \emph{causal}. Question 2 asks if the program is the \emph{cause} of the lower rearrest rate among the participants. In other words, it asks about the \emph{mechanism} behind the data. Question 3 asks a prediction of the rearrest rate \emph{after some novel outside intervention to the system}, namely after making the program compulsory for all prisoners. In order to make such a prediction, one needs to understand the causal structure of the system.

\begin{example}\label{ex: genes}
   We consider gene expression levels of yeast cells. Suppose that we want to predict the average gene expression levels after knocking out one of the genes, or after knocking out multiple genes at a time. These are again causal questions, since we want to make predictions after interventions to the system.
%
\end{example}

Thus, causal questions are about the \emph{mechanism} behind the data or about predictions \emph{after a novel intervention is applied to the system}. They arise in all parts of science. Application areas involving big data include for example systems biology (e.g., \cite{ChuEtAl03, FriedmanEtAl00, MaEtAl14, MaathuisColomboKalischBuehlmann10, Opgen-RheinStrimmer07, StekhovenEtAl12}), neuroscience (e.g., \cite{ChicharroPanzeri14,HansenEtAl13,RamseyEtAl10,SmithEtAl11}), climate science (e.g., \cite{Ebert-UphoffDeng12, Ebert-UphoffDeng15}), and marketing (e.g., \cite{BrodersenEtAl15}).


\subsection{Observational versus experimental data}\label{sec: obs versus exp data}

Going back to the prisoners example, which of the three posed questions can we answer? This depends on the origin of the data, and brings us to the distinction between observational and experimental data.\\

\noindent{\bf Observational data}.
Suppose first that participation in the program was voluntary. Then we would have so-called \emph{observational data}, since the subjects (prisoners) chose their own treatment (rehabilitation program or not), while the researchers just \emph{observed} the results. From observational data, we can easily answer question 1. It is difficult, however, to answer questions 2 and 3.

Let us first consider question 2. Since the participants form a self-selected subgroup, there may be many differences between the participants and the non-participants. For example, the participants may be more motivated to change their lives, and this may contribute to the difference in rearrest rates. In this case, the effects of the program and the motivation of the prisoners are said to be mixed-up or \emph{confounded}.

Next, let us consider question 3. At first sight, one may think that the answer is simply $20\%$, since this was the rearrest rate among the participants of the program. But again we have to keep in mind that the participants form a self-selected subgroup that is likely to have special characteristics. Hence, the rearrest rate of this subgroup cannot be extrapolated to the entire prisoners population. \\

\noindent{\bf Experimental data}.
Now suppose that it was up to the researchers to decide which prisoners participated in the program. For example, suppose that the researchers rolled a die for each prisoner, and let him/her participate if the outcome was 1 or 2. Then we would have a so-called \emph{randomized controlled experiment} and \emph{experimental data}.

Let us look again at question 2. Due to the randomization, the motivation level of the prisoners is likely to be similar in the two groups. Moreover, any other factors of importance (like social background, type of crime committed, number of earlier crimes, etcetera) are likely to be similar in the two groups. Hence, the groups are equal in all respects, except for participation in the program. The observed difference in rearrest rate must therefore be due to the program. This answers question 2.

Finally, the answer to question 3 is now $20\%$, since the randomized treatment assignment ensures that the participants form a representative sample of the population. \\

Thus, causal questions are best answered by experimental data, and we should work with such data whenever possible. Experimental data is not always available, however, since randomized controlled experiments can be unethical, infeasible, time consuming or expensive.
On the other hand, observational data is often relatively cheap and abundant. In this paper, we therefore consider the problem of answering causal questions about large-scale systems from observational data.

\subsection{Problem formulation}\label{sec: outline}

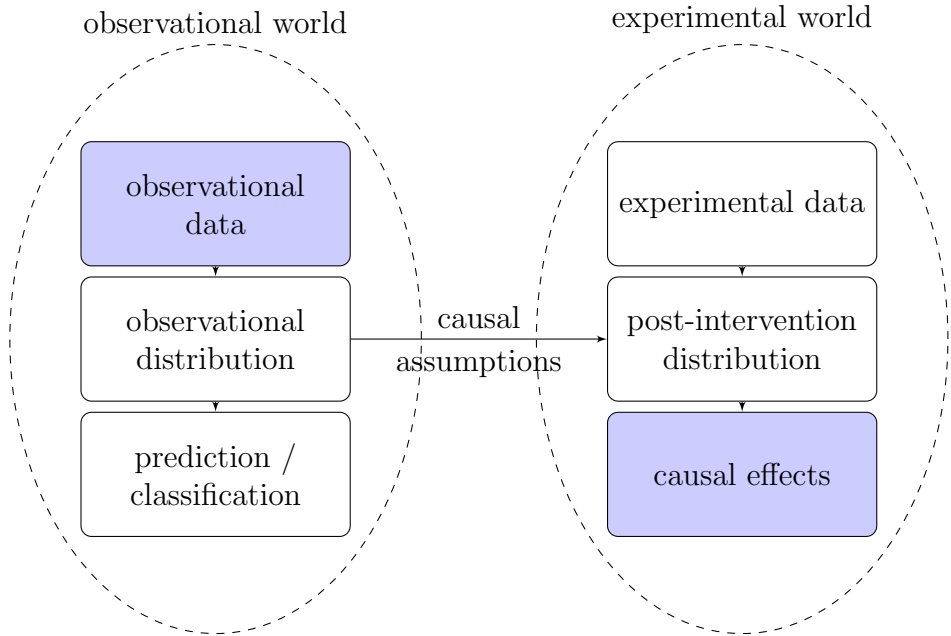
\begin{figure}[t,h]
\centering
\tikzstyle{block} = [rectangle, draw,
    text width=8em, text centered, rounded corners, minimum height=4em]
\tikzstyle{line} = [draw, -latex']
\begin{centering}
\begin{tikzpicture}[node distance = 1.8cm, auto]
    \node [block, fill=blue!20] (obsdata) {observational data};
    \node [block, right of=obsdata, node distance=7cm] (expdata) {experimental data};
    \node [block, below of=obsdata] (obsdist) {observational distribution};
    \node [block, below of=expdata] (postintdist) {post-intervention distribution};
    \node [block, below of=obsdist] (pred) {prediction / classification};
    \node [block, below of=postintdist, fill=blue!20] (causal) {causal effects};
    \node[shape=ellipse, draw, dashed, fit=(obsdata)(obsdist)(pred), label={observational world}] (obsworld) {};
    \node[shape=ellipse, draw, dashed, fit=(expdata)(postintdist)(causal), label={experimental world}] (expworld) {};
    \path [line] (obsdata) -> (obsdist);
    \path [line] (obsdist) -> (pred);
    \path [line] (expdata) -> (postintdist);
    \path [line] (postintdist) -> (causal);
    \draw [line] (obsdist) -> node[above] {causal} node[below] {assumptions} ++ (postintdist);
\end{tikzpicture}
\end{centering}
\caption{We want to estimate causal effects from observational data. This means that we need to move from the observational world to the experimental world. This can only done by imposing causal assumptions.}
\label{fig: obs-exp worlds}
\end{figure}

It is relatively straightforward to make ``standard" predictions based on observational data (see the ``observational world" in Figure \ref{fig: obs-exp worlds}), or to estimate causal effects from randomized controlled experiments (see the ``experimental world" in Figure \ref{fig: obs-exp worlds}). But we want to estimate \emph{causal effects} from \emph{observational data}.
%
This means that we need to move from the observational world to the experimental world. This step is fundamentally impossible without causal assumptions, even in the large sample limit with perfect knowledge about the observational distribution (cf. Section 2 of \cite{Pearl09}). In other words, causal assumptions are needed to deduce the post-intervention distribution from the observational distribution. In this paper, we assume that the data were generated from a (known or unknown) causal structure which can be represented by a directed acyclic graph (DAG). \\


\noindent{\bf Outline of this paper.} In the next section, we assume that the data were generated from a known DAG. In particular, we discuss the framework of a structural equation model (SEM) and its corresponding causal DAG. We also discuss the estimation of causal effects under such a model. In large-scale networks, however, the causal DAG is often unknown. Next, we therefore discuss causal structure learning, that is, learning information about the causal structure from observational data. We then combine these two parts and discuss methods to estimate
(bounds on) causal effects from observational data when the causal structure is unknown. We also illustrate this method on a yeast gene expression data set. We close by mentioning several extensions of the discussed work.

\section{Estimating causal effects when the causal structure is known}\label{sec: estimating causal effects when DAG is known}


Causal structures can be represented by graphs, where the random variables are represented by nodes (or vertices), and causal relationships between the variables are represented by edges between the corresponding nodes. Such causal graphs have two important practical advantages. First, a causal graph provides a transparent and compact description of the causal assumptions that are being made. This allows these assumptions to be discussed and debated among researchers. Next, after agreeing on a causal graph, one can easily determine causal effects. In particular, we can read off from the graph which sets of variables can or cannot be used for covariate adjustment in order to obtain a given causal effect. We refer to \cite{Pearl09,Pearl09book} for further details on the material in this section.


\subsection{Graph terminology}\label{sec: graph terminology}

We consider graphs with \emph{directed edges} ($\to$) and \emph{undirected edges} ($-$). There can be at most one edge between any pair of distinct nodes. If all edges are directed (undirected), then the graph is called \emph{directed} (\emph{undirected}). A \emph{partially directed graph} can contain both directed and undirected edges. The \emph{skeleton} of a partially directed graph is the undirected graph that results from replacing all directed edges by undirected edges.

Two nodes are \emph{adjacent} if they are connected by an edge. If $X\to Y$, then $X$ is a parent of $Y$. The adjacency set and the parent set of a node $X$ in a graph ${\cal G}$ are denoted by $adj(X,{\cal G})$ and $pa(X, {\cal G})$,  respectively. A graph is \emph{complete} if every pair of nodes is adjacent.

A \emph{path} in a graph ${\cal G}$ is a distinct sequence of nodes, such that all successive pairs of nodes in the sequence are adjacent in ${\cal G}$. A \emph{directed path} from $X$ to $Y$ is a path between $X$ and $Y$ in which all edges point towards $Y$, i.e., $X \to \dots \to Y$. A directed path from $X$ to $Y$ together with an edge $Y \to X$ forms a \emph{directed cycle}. A directed graph is \emph{acyclic} if it does not contain directed cycles. A \emph{directed acyclic graph} is also called a \emph{DAG}.

A node $X$ is a \emph{collider} on a path if the path has two colliding arrows at $X$, that is, the path contains $ \to X \leftarrow $. Otherwise $X$ is a \emph{non-collider} on the path. We emphasize that the collider status of a node is relative to a path; a node can be a collider on one path, while it is a non-collider on another. The collider $X$ is \emph{unshielded} if the neighbors of $X$ on the path are not adjacent to each other in the graph, that is, the path contains $W \to X \leftarrow Z$ and $W$ and $Z$ are not adjacent in the graph.

\subsection{Structural equation model (SEM)}\label{sec: structural causal model}

We consider a collection of random variables $X_1,\dots, X_p$ that are generated by structural equations (see, e.g. \cite{Bollen89, Wright21}):
\begin{align}\label{eq: structural causal model}
   X_i \leftarrow g_i({\bf S}_i, \epsilon_i), \quad \qquad  i=1,\dots,p,
\end{align}
where ${\bf S}_i \subseteq \{X_1,\dots,X_p\}\setminus \{X_i\}$ and $\epsilon_i$ is some random noise. We interpret these equations causally, as describing how each $X_i$ is generated from the variables in ${\bf S}_i$ and the noise $\epsilon_i$. Thus, changes to the variables in $\mathbf{S}_i$ can lead to changes in $X_i$, but not the other way around. We use the notation $\leftarrow$ in \eqref{eq: structural causal model} to emphasize this asymmetric relationship. Moreover, we assume that the structural equations are \emph{autonomous}, \label{ass: autonomy} in the sense that we can change one structural equation without affecting the others. This will allow the modelling of local interventions to the system.

The structural equations correspond to a directed graph $\cal{G}$ that is generated as follows: the nodes are given by $X_1,\dots,X_p$, and the edges are drawn so that ${\bf S}_i$ is the parent set of $X_i$, $i=1,\dots,p$. The graph $\cal G$ then describes the causal structure and is called the \emph{causal graph}: the presence of an edge $X_j \to X_i$ means that $X_j$ is a potential direct cause of $X_i$ (i.e., $X_j$ may play a role in the generating mechanism of $X_i$), and the absence of an edge $X_k \to X_i$ means that $X_k$ is definitely not a direct cause of $X_i$ (i.e., $X_k$ does not play a role in the generating mechanism of $X_i$).

Throughout, we make several assumptions about the model. The graph ${\cal G}$ is assumed to be acyclic (hence a DAG), and the error terms $\epsilon_1,\dots,\epsilon_p$ are jointly independent. In terms of the causal interpretation, these assumptions mean that we do not allow feedback loops nor unmeasured confounding variables. The above model with these assumptions was called a \emph{structural causal model} by \cite{Pearl95}. We will simply refer to it as a \emph{structural equation model (SEM)}. If all structural equations are linear, we will call it a \emph{linear SEM}.\\

We now discuss two important properties of SEMs, namely factorization and d-separation.
If $X_1,\dots,X_p$ are generated from a SEM with causal DAG $\cal G$, then the density $f(x_1,\dots,x_p)$ of $X_1,\dots,X_p$ (assuming it exists) factorizes as:
\begin{align}\label{eq: factorization}
   f(x_1,\dots,x_p) = \prod_{i=1}^p f_i(x_i|pa(x_i,\cal G)),
\end{align}
where $f_i(x_i|pa(x_i,\cal G))$ is the conditional density of $X_i$ given $pa(X_i,\cal G)$.

If a density factorizes according to a DAG as in \eqref{eq: factorization}, then one can use the DAG to read off conditional independencies that must hold in the distribution (regardless of the choice of the $f_i(\cdot)$'s), using a graphical criterion called \emph{d-separation} (see, e.g., Definition 1 in \cite{Pearl09}). In particular, the so-called global Markov property implies that when two disjoint sets $\mathbf{A}$ and $\mathbf{B}$ of vertices are d-separated by a third disjoint set $\mathbf{S}$, then $\mathbf{A}$ and $\mathbf{B}$ are conditionally independent given $\mathbf{S}$ ($\mathbf{A}\ci \mathbf{B}|\mathbf{S}$) in any distribution that factorizes according to the DAG.

\begin{example}\label{ex: structural causal model}
  We consider the following structural equations and the corresponding causal DAG for the random variables $P$, $S$, $R$ and $M$:\\
  \begin{minipage}{0.5\textwidth}
    \begin{align*}
       P & \leftarrow g_1(M, \epsilon_P)\\
      S & \leftarrow g_2(P, \epsilon_S)\\
      R & \leftarrow g_3(M, S, \epsilon_R)\\
       M & \leftarrow g_4(\epsilon_M)
    \end{align*}
    \vspace{.15cm}
  \end{minipage}
  \begin{minipage}{0.5\textwidth}
       \begin{tikzpicture}[>=stealth',shorten >=1pt,auto,node distance=2cm,main node/.style={minimum size=0.6cm,font=\sffamily\Large\bfseries},scale=0.6,transform shape]
   \node[main node]         (X1) at (6,0)                      {$P$};
   \node[main node]         (X2) [right of =X1]  	{$S$};
   \node[main node]         (X3) [right of= X2]  	{$R$};
   \node[main node]       	(X4) [above of= X2]         	{$M$};
   \draw[->] (X4) edge    (X1);
   \draw[->] (X4) edge    (X3);
   \draw[->] (X1) edge    (X2);
   \draw[->] (X2) edge    (X3);
   \end{tikzpicture}
  \end{minipage}
  where $\epsilon_P$, $\epsilon_S$, $\epsilon_R$ and $\epsilon_M$ are mutually independent with arbitrary mean zero distributions.
  For each structural equation, the variables on the right hand side appear in the causal DAG as the parents of the variable on the left hand side.

   We denote the random variables by $M$, $P$, $S$ and $R$, since these structural equations can be used to describe a possible causal mechanism behind the prisoners data (Example \ref{ex: prisoners}), where $M=\, $measure of motivation, $P=\, $participation in the program ($P=1$ means participation, $P=0$ otherwise), $S=\, $measure of social skills taught by the program, and $R=\, $rearrest ($R=1$ means rearrest, $R=0$ otherwise).

   We see that the causal DAG of this SEM indeed provides a clear and compact description its causal assumptions. In particular, it allows that motivation directly affects participation and rearrest. Moreover, it allows that participation directly affects social skills, and that social skills directly affect rearrest. The missing edge between $M$ and $S$ encodes the assumption that there is no direct effect from motivation on social skills. In other words, any effect of motivation on social skills goes entirely through participation (see the path $M\to P \to S$). Similarly, the missing edge between $P$ and $R$ encodes the assumption that there is no direct effect of participation on rearrest; any effect of participation on rearrest must fully go through social skills (see the path $P \to S \to R$).
\end{example}

\subsection{Post-intervention distributions and causal effects}\label{sec: determining post-intervention distribution}

Now how does the framework of the SEM allow us to move between the observational and experimental worlds?
This is straightforward, since an intervention at some variable $X_i$ simply means that we change the generating mechanism of $X_i$, that is, we change the corresponding structural equation $g_i(\cdot)$ (and leave the other structural equations unchanged). For example, one can let $X_i  \leftarrow \epsilon_i$ where $\epsilon_i$ has some given distribution, or $X_i \leftarrow x_i'$ for some fixed value $x_i'$ in the support of $X_i$. The latter is often denoted as Pearl's do-intervention $do(X_i=x_i')$ and is interpreted as setting the variable $X_i$ to the value $x_i'$ by an outside intervention, uniformly over the entire population \cite{Pearl09}.

\begin{example}\label{ex: do-operator}
  In the prisoners example (see Examples \ref{ex: prisoners} and \ref{ex: structural causal model}), the quantity $P(R=1|do(P=1))$ represents the rearrest probability when all prisoners are forced to participate in the program, while $P(R=1|do(P=0))$ is the rearrest probability if no prisoner is allowed to participate in the program. We emphasize that these quantities are generally not equal to the usual conditional probabilities $P(R=1|P=1)$ and $P(R=1|P=0)$, which represent the rearrest probabilities among prisoners who choose to participate or not to participate in the program.

  In the gene expression example (see Example \ref{ex: genes}), let $X_i$ and $X_j$ represent the gene expression level of genes $i$ and $j$. Then $E(X_j | do(X_i=x_i'))$ represents the average expression level of gene $j$ after setting the gene expression level of gene $i$ to the value $x_i'$ by an outside intervention.
\end{example}

\noindent{\bf Truncated factorization formula.} A do-intervention on $X_i$ means that $X_i$ no longer depends on its former parents in the DAG, so that the incoming edges into $X_i$ can be removed. This leads to a so-called truncated DAG. The post-intervention distribution factorizes according to this truncated DAG, so that we get:
\begin{align}\label{eq: truncated factorization}
   f(x_1,\dots,x_p|do(X_i=x_i')) = \left\{ \begin{array}{ll}
                                \prod_{j\neq i} f_j(x_j|pa(x_j,\cal G)) & \text{if}\, x_i=x_i',\\
                                0  & \text{otherwise.}
                                \end{array}\right.
\end{align}
This is called the truncated factorization formula \cite{Pearl93}, the manipulation formula \cite{SpirtesEtAl93} or the g-formula \cite{Robins86}. Note that this formula heavily uses the factorization formula \eqref{eq: factorization} and the ``autonomy assumption" (see page \pageref{ass: autonomy}).\\

\noindent{\bf Defining the total effect.}
Summary measures of the post-intervention distribution can be used to define total causal effects. In the prisoners example, it is natural to define the total effect of $P$ on $R$ as  $$P(R=1|do(P=1))-P(R=1|do(P=0)).$$ Again, we emphasize that this is different from $P(R=1|P=1)-P(R=1|P=0)$.

In a setting with continuous variables, the total effect of $X_i$ on $Y$ can be defined as $$\left. \frac{\partial}{\partial x_i} E(Y|do(X_i = x_i)\right|_{x_i=x_i'}.$$\\

\noindent{\bf Computing the total effect.}
 A total effect can be computed using, for example, covariate adjustment \cite{Pearl09, ShpitserVanderWeeleRobins10}, inverse probability weighting (IPW) \cite{RobinsHernanBrumback00,HernanEtAl00}, or instrumental variables (e.g, \cite{AngristEtAl96}). In all these methods, the causal DAG plays an important role, since it tells us which variables can be used for covariate adjustment, which variables can be used as instruments, or which weights should be used in IPW.

 In this paper, we focus mostly on linear SEMs. In this setting, the total effect of $X_i$ on $Y$ can be easily computed via linear regression with covariate adjustment. If $Y \in pa(X_i,\cal G)$ then the effect of $X_i$ on $Y$ equals zero. Otherwise, it equals the regression coefficient of $X_i$ in the linear regression of $Y$ on $X_i$ and $pa(X_i,\cal{G})$ (see Proposition 3.1 of \cite{NandyMaathuisRichardson14b}). In other words, we simply regress $Y$ on $X_i$ while adjusting for the parents of $X_i$ in the causal DAG. This is also called ``adjusting for direct causes of the intervention variable".

\begin{example}\label{ex: linear SEM}
  We consider the following linear SEM:\\
    \begin{minipage}{0.5\textwidth}
    \begin{align*}
       X_1 & \leftarrow 2 X_4 +  \epsilon_1\\
      X_2 & \leftarrow 3 X_1 + \epsilon_2\\
      X_3 & \leftarrow  2 X_2  + X_4 + \epsilon_3\\
       X_4 & \leftarrow \epsilon_4
    \end{align*}
    \vspace{.15cm}
  \end{minipage}
  \begin{minipage}{0.5\textwidth}
       \begin{tikzpicture}[>=stealth',shorten >=1pt,auto,node distance=2cm,main node/.style={minimum size=0.6cm,font=\sffamily\Large\bfseries},scale=0.6,transform shape]
   \node[main node]         (X1) at (6,0)                      {$X_1$};
   \node[main node]         (X2) [right of =X1]  	{$X_2$};
   \node[main node]         (X3) [right of= X2]  	{$X_3$};
   \node[main node]       	(X4) [above of= X2]         	{$X_4$};
   \draw[->] (X4) edge  node[above=5pt]{\Large{$2$}}  (X1);
   \draw[->] (X4) edge  node{\Large{$1$}} (X3);
   \draw[->] (X1) edge   node[below=2pt]{\Large{$3$}} (X2);
   \draw[->] (X2) edge   node[below=2pt]{\Large{$2$}} (X3);
   \end{tikzpicture}
  \end{minipage}.
  The errors are mutually independent with arbitrary mean zero distributions. We note that the coefficients in the structural equations are depicted as edge weights in the causal DAG.

  Suppose we are interested in the total effect of $X_1$ on $X_3$. Then we consider an outside intervention that sets $X_1$ to the value $x_1$, i.e., $do(X_1=x_1)$. This means that we change the structural equation for $X_1$ to $X_1 \leftarrow x_1$.  Since the other structural equations do not change, we then obtain $X_2= 3x_1+\epsilon_2$, $X_4 = \epsilon_4$ and $X_3 = 2X_2 + X_4 + \epsilon_3 = 6x_1 + 2\epsilon_2 + \epsilon_4 + \epsilon_3$. Hence, $E(X_3 | do(X_1=x_1)) = 6x_1$, and differentiating  with respect to $x_1$ yields a total effect of $6$.

  We note that the total effect of $X_1$ on $X_3$ also equals the product of the edge weights along the directed path $X_1 \to X_2 \to X_3$. This is true in general for linear SEMs: the total effect of $X_i$ on $Y$ can be obtained by multiplying the edge weights along each directed path from $X_i$ to $Y$, and then summing over the directed paths (if there is more than one).

  The total effect can also be obtained via regression. Since $pa(X_1,\mathcal G) = \{X_4\}$, the total effect of $X_1$ on $X_3$ equals the coefficient of $X_1$ in the regression of $X_3$ on $X_1$ and $X_4$. It can be easily verified that this again yields 6. One can also verify that adjusting for any other subset of $\{X_2,X_4\}$ does not yield the correct total effect.
\end{example}

\section{Causal structure learning}\label{sec: causal structure learning}

The material in the previous section can be used if the causal DAG is known. In settings with big data, however, it is rare that one can draw the causal DAG. In this section, we therefore consider methods for learning DAGs from observational data. Such methods are called \emph{causal structure learning methods}.

Recall from Section \ref{sec: structural causal model} that DAGs encode conditional independencies via d-separation. Thus, by considering conditional independencies in the observational distribution, one may hope to reverse-engineer the causal DAG that generated the data. Unfortunately, this does not work in general, since the same set of d-separation relationships can be encoded by several DAGs. Such DAGs are called \emph{Markov equivalent} and form a \emph{Markov equivalence class}.

A Markov equivalence class can be described uniquely by a completed partially directed acyclic graph (CPDAG) \cite{AnderssonEtAl97, Chickering02}. The skeleton of the CPDAG is defined as follows. Two nodes $X_i$ and $X_j$ are adjacent in the CPDAG if and only if, in any DAG in the Markov equivalence class, $X_i$ and $X_j$ cannot be d-separated by any set of the remaining nodes. The orientation of the edges in the CPDAG is as follows. A directed edge $X_i \to X_j$ in the CPDAG means that the edge $X_i \to X_j$ occurs in all DAGs in the Markov equivalence class. An undirected edge $X_i - X_j$ in the CPDAG means that there is a DAG in the Markov equivalence class with $X_i \to X_j$, as well as a DAG with $X_i \leftarrow X_j$.

It can happen that a distribution contains more conditional independence relationships than those that are encoded by the DAG via d-separation. If this is \emph{not} the case, then the distribution is called \emph{faithful} with respect to the DAG. If a distribution is both Markov and faithful with respect to a DAG, then the conditional independencies in the distribution correspond exactly to d-separation relationships in the DAG, and the DAG is called a \emph{perfect map} of the distribution.\\

\noindent{\bf Problem setting.} Throughout this section, we consider the following setting. We are given $n$ i.i.d.\ observations of $\mathbf{X}$, where $\mathbf{X}=(X_1,\dots,X_p)$ is generated from a SEM. We assume that the corresponding causal DAG $\cal G$ is a perfect map of the distribution of $\mathbf{X}$. We aim to learn the Markov equivalence class of $\cal G$.\\

In the following three subsections we discuss so-called constraint-based, score-based and hybrid methods for this task. The discussed algorithms are available in the R-package \texttt{pcalg} \cite{KalischEtAl12}. In the last subsection we discuss a class of methods that can be used if one is willing to impose additional restrictions on the SEM that allow identification of the causal DAG (rather than its CPDAG).

\subsection{Constraint-based methods}\label{sec: PC}

Constraint-based methods learn the CPDAG by exploiting conditional independence constraints in the observational distribution. The most prominent example of such a method is probably the PC algorithm \cite{SpirtesEtAl00}. This algorithm first estimates the skeleton of the underlying CPDAG, and then determines the orientation of as many edges as possible.

We discuss the estimation of the skeleton in more detail. Recall that, under the Markov and faithfulness assumptions, two nodes $X_i$ and $X_j$ are adjacent in the CPDAG if and only if they are conditionally dependent given all subsets of $\mathbf{X} \setminus \{X_i,X_j\}$. Therefore, adjacency of $X_i$ and $X_j$ can be determined by testing $X_i \ci X_j |\mathbf{S}$ for all possible subsets $\mathbf{S} \subseteq \mathbf{X} \setminus \{X_i,X_j\}$. This naive approach is used in the SGS algorithm \cite{SpirtesEtAl00}. It quickly becomes computationally infeasible for a large number of variables.

The PC algorithm avoids this computational trap by using the following fact about DAGs: two nodes $X_i$ and $X_j$ in a DAG $\cal G$ are d-separated by some subset of the remaining nodes if and only if they are d-separated by $pa(X_i,\cal G)$ or by $pa(X_j, \cal G)$. This fact may seem of little help at first, since we do not know $pa(X_i,\cal G)$ and $pa(X_j,\cal G)$ (then we would know the DAG!). It is helpful, however, since it allows a clever ordering of the conditional independence tests in the PC algorithm, as follows. The algorithm starts with a complete undirected graph. It then assesses, for all pairs of variables, whether they are marginally independent. If a pair of variables is found to be independent, then the edge between them is removed.  Next, for each pair of nodes $(X_i,X_j)$ that are still adjacent, it tests conditional independence of the corresponding random variables given all possible subsets of size 1 of $adj(X_i,{\cal G}^*)\setminus\{X_j\}$ and of $adj(X_j,{\cal G}^*)\setminus\{X_i\}$, where $\cal G^*$ is the current graph. Again, it removes the edge if such a conditional independence is deemed to be true. The algorithm continues in this way, considering conditioning sets of increasing size, until the size of the conditioning sets is larger than the size of the adjacency sets of the nodes.

This procedure gives the correct skeleton when using perfect conditional independence information. To see this, note that at any point in the procedure, the current graph is a supergraph of the skeleton of the CPDAG. By construction of the algorithm, this ensures that $X_i \ci X_j |pa(X_i,\cal G)$ and $X_i\ci X_j |pa(X_j,\cal G)$ were assessed.

After applying certain edge orientation rules, the output of the PC algorithm is a partially directed graph, the estimated CPDAG. This output depends on the ordering of the variables (except in the limit of an infinite sample size), since the ordering determines which conditional independence tests are done. This issue was studied in \cite{ColomboMaathuis14}, where it was shown that the order-dependence can be very severe in high-dimensional settings with many variables and a small sample size (see Section \ref{sec: validations IDA} for a data example). Moreover, \cite{ColomboMaathuis14} proposed an order-independent version of the PC algorithm, called PC-stable. This version is now the default implementation in the R-package \texttt{pcalg} \cite{KalischEtAl12}.


We note that the user has to specify a significance level $\alpha$ for the conditional independence tests. Due to multiple testing, this parameter does \emph{not} play the role of an overall significance level. It should rather be viewed as a tuning parameter for the algorithm, where smaller values of $\alpha$ typically lead to sparser graphs.

%

The PC and PC-stable algorithms are computationally feasible for sparse graphs with thousands of variables.
Both PC and PC-stable were shown to be consistent in sparse high-dimensional settings, when the joint distribution is multivariate Gaussian and conditional independence is assessed by testing for zero partial correlation  \cite{KalischBuehlmann07a,ColomboMaathuis14}. By using Rank correlation, consistency can be achieved in sparse high-dimensional settings for a broader class of Gaussian copula or nonparanormal models \cite{HarrisDrton13}.




\subsection{Score-based methods}\label{sec: GES}

Score-based methods learn the CPDAG by (greedily) searching for an optimally scoring DAG, where the score measures how well the data fits to the DAG, while penalizing the complexity of the DAG.

A prominent example of such an algorithm is the greedy equivalence search (GES) algorithm \cite{Chickering03}.
GES is a grow-shrink algorithm that consists of two phases: a forward phase and a backward phase. The forward phase starts with an initial estimate (often the empty graph) of the CPDAG, and sequentially adds single edges, each time choosing the edge addition that yields the maximum improvement of the score, until the score can no longer be improved. The backward phase starts with the output of the forward phase, and sequentially deletes single edges, each time choosing the edge deletion that yields a maximum improvement of the score, until the score can no longer be improved. A computational advantage of GES over the traditional DAG-search methods is that it searches over the space of all possible CPDAGs, instead of over the space of all possible DAGs.

The GES algorithm requires the scoring criterion to be \emph{score equivalent}, meaning that every DAG in a Markov equivalence class gets the same score. Moreover, the choice of scoring criterion is crucial for computational and statistical performances. The so-called \emph{decomposability} property of a scoring criterion allows fast updates of scores during the forward and the backward phase. For example, (penalized) log-likelihood scores are decomposable, since \eqref{eq: factorization} implies that the (penalized) log-likelihood score of a DAG can be computed by summing up the (local) scores of each node given its parents in the DAG. Finally, the so-called \emph{consistency} property of a scoring criterion ensures that the true CPDAG gets the highest score with probability approaching one (as the sample size tends to infinity).

GES was shown to be consistent when the scoring criterion is score equivalent, decomposable and consistent. For multivariate Gaussian or multinomial distributions, penalized likelihood scores such as BIC satisfy these assumptions.

\subsection{Hybrid methods}\label{sec: hybrid}

Hybrid methods learn the CPDAG by combining the ideas of constraint-based and score-based methods. Typically, they first estimate (a supergraph of) the skeleton of the CPDAG using conditional independence tests, and then apply a search and score technique while restricting the set of allowed edges to the estimated skeleton. A prominent example is the Max-Min Hill-Climbing (MMHC) algorithm \cite{TsamardinosEtAl06}.


The restriction on the search space of hybrid methods provides a huge computational advantage when the estimated skeleton is sparse. This is why the hybrid methods scale well to thousands of variables, whereas the unrestricted score-based methods do not. However, this comes at the cost of inconsistency or at least at the cost of a lack of consistency proofs. Interestingly, empirical results have shown that a restriction on the search space can also help to improve the estimation quality \cite{TsamardinosEtAl06}.

This gap between theory and practice was addressed in \cite{NandyEtAl15}, who proposed a consistent hybrid modification of GES, called ARGES. The search space of ARGES mainly depends on an estimated conditional independence graph. (This is an undirected graph containing an edge between $X_i$ and $X_j$ if and only if $X_i \nci X_j | \mathbf{V}\setminus\{X_i,X_j\}$. It is a  supergraph of the skeleton of the CPDAG.) But the search space also changes adaptively depending on the current state of the algorithm. This adaptive modification is necessary to achieve consistency in general. The fact that the modification is relatively minor may provide an explanation for the empirical success of (inconsistent) hybrid methods.

\subsection{Learning SEMs with additional restrictions}\label{sec: lingam}

Now that we have looked at various different methods to estimate the CPDAG, we close this section by discussing a slightly different approach that allows estimation of the causal DAG rather than its CPDAG. Identification of the DAG can be achieved by imposing additional restrictions on the generating SEM. Examples of this approach include the LiNGAM method for linear SEMs with non-Gaussian noise \cite{ShimizuEtAl06-JMLR, ShimizuEtAl09}, methods for nonlinear SEMs \cite{HoyerEtAl09} and methods for linear Gaussian SEMs with equal error variances \cite{PetersBuhlmann14}.

We discuss the LiNGAM method in some more detail. A linear SEM can be written as $\mathbf{X} = B \mathbf{X} + \boldsymbol\epsilon$ or equivalently $\mathbf{X} = A \boldsymbol\epsilon$ with $A = (I-B)^{-1}$. The LiNGAM algorithm of \cite{ShimizuEtAl06-JMLR} uses independent component analysis (ICA) to obtain estimates $\hat{A}$ and $\hat{B} = I - \hat{A}^{-1}$ of $A$ and $B$. Ideally, rows and columns of $\hat{B}$ can be permuted to obtain a lower triangular matrix and hence an estimate of the causal DAG. This is not possible in general in the presence of sampling errors, but a lower triangular matrix can be obtained by setting some small non-zero entries to zero and permuting rows and columns of $\hat{B}$.

A more recent implementation of the LiNGAM algorithm, called DirectLiNGAM was proposed by \cite{ShimizuEtAl09}. This implementation is not based on ICA. Rather, it estimates the variable ordering by iteratively finding
an exogenous variable. DirectLiNGAM is suitable for settings with a larger number of variables.



\section{Estimating the size of causal effects when the causal structure is unknown}\label{sec: estimating causal effects when DAG is unknown}

We now combine the previous two sections and discuss methods to estimate bounds on causal effects from observational data when the causal structure is unknown. We first define the problem setting. \\

\noindent{\bf Problem setting:} We have $n$ i.i.d.\ realizations of $\mathbf{X}$, where $\mathbf{X}$ is generated from a linear SEM with Gaussian errors. We do not know the corresponding causal DAG, but we assume that it is a perfect map of the distribution of $\mathbf{X}$. Our goal is to estimate the sizes of causal effects. \\

We first discuss the IDA method \cite{MaathuisKalischBuehlmann09} to estimate the effect of single interventions in this setting (for example a single gene knockout). Next, we consider a generalization of this approach for multiple simultaneous interventions, called jointIDA \cite{NandyMaathuisRichardson14b}. Finally, we present a data application from \cite{MaathuisColomboKalischBuehlmann10,ColomboMaathuis14}.


\subsection{IDA}\label{sec: IDA}

Suppose we want to estimate the total effect of $X_1$ on a response variable $Y$. The conceptual idea of IDA is as follows. We first estimate the CPDAG of the underlying causal DAG, using for example the PC algorithm. Next, we can list all the DAGs in the Markov equivalence class described by the estimated CPDAG. Under our assumptions and in the large sample limit, one of these DAGs is the true causal DAG. We can then apply covariate adjustment for each DAG, yielding an estimated total effect of $X_1$ on $Y$ for each possible DAG. We collect all these effects in a multiset $\hat \Theta$. Bounds on $\hat \Theta$ are estimated bounds on the true causal effect.

For large graphs, it is computationally intensive to list all the DAGs in the Markov equivalence class. However, since we can always use the parent set of $X_1$ as adjustment set (see Section \ref{sec: determining post-intervention distribution}), it suffices to know the parent set of $X_1$ for each of the DAGs in the Markov equivalence class, rather than the entire DAGs. These possible parent sets of $X_1$ can be extracted easily from the CPDAG. It is then only left to count the number of DAGs in the Markov equivalence class with each of these parent sets. In \cite{MaathuisKalischBuehlmann09} the authors used a shortcut, where they only looked whether a parent set is locally valid or not, instead of counting the number of DAGs in the Markov equivalence class. Here locally valid means that the parent set does not create a new unshielded collider with $X_1$ as collider. This shortcut results in a set $\hat \Theta^L$ which contains the same distinct values as $\hat \Theta$, but might have different multiplicities. Hence, if one is only interested in bounds on causal effects, the information in $\hat \Theta^L$ is sufficient. In other cases, however, the information on multiplicities might be important, for example if one is interested in the direction of the total effect ($\hat \Theta = \{1,1,1,1,1, -1\}$ would make us guess the effect is positive, while $\hat \Theta^L = \{1, -1\}$ loses this information).

IDA was shown to be consistent in sparse high-dimensional settings.

\begin{figure}
   \begin{center}
      \includegraphics[width=0.9\textwidth]{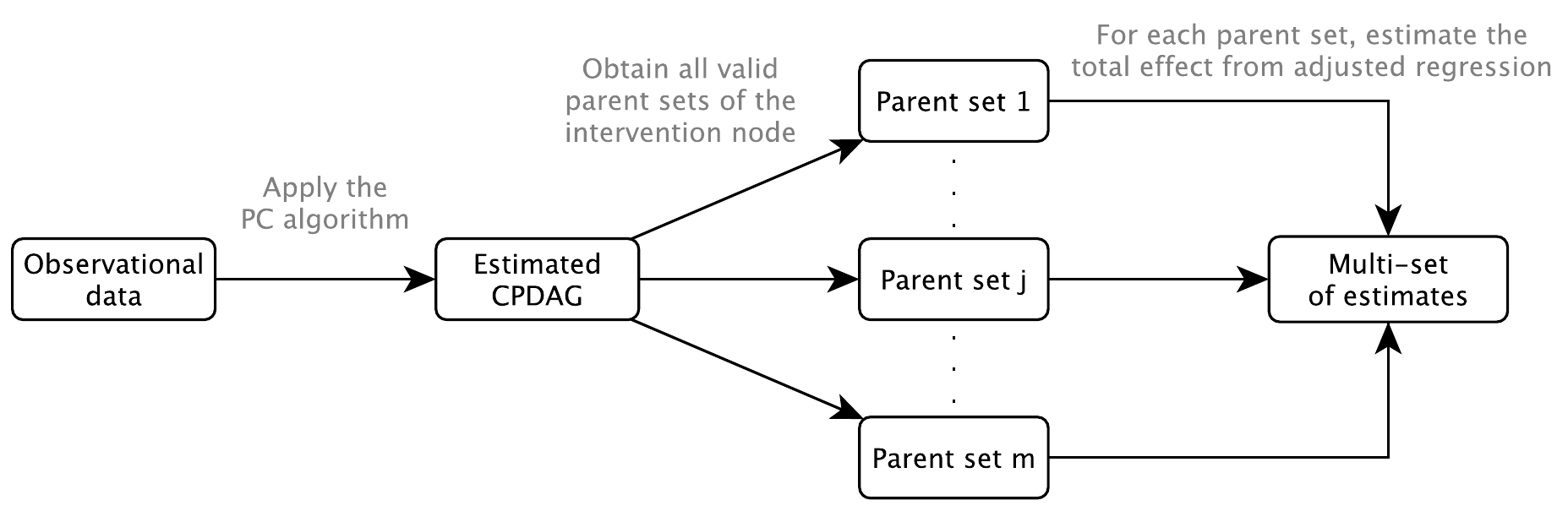}
      \end{center}
      \label{ida}
      \caption{Schematic representation of the IDA algorithm, taken from \cite{NandyMaathuisRichardson14b}.}
\end{figure}

\subsection{JointIDA}\label{sec: jointIDA}

We can also estimate the effect of multiple simultaneous or joint interventions. For example, we may want to predict the effect of a double or triple gene knockout.

Generalizing IDA to this setting poses several non-trivial challenges. First, even if the parent sets of the intervention sets are known, it is non-trivial to estimate the size of a total joint effect, since a straightforward adjusted regression no longer works. Available methods for this purpose are IPW \cite{RobinsHernanBrumback00} and the recently developed methods RRC \cite{NandyMaathuisRichardson14b} and MCD \cite{NandyMaathuisRichardson14b}.
Under our assumptions, RRC recursively computes joint effects from single intervention effects, and MCD produces an estimate of the covariance matrix of the interventional distribution by iteratively modifying Cholesky decompositions of covariance matrices.

Second, we must extract possible parent sets for the intervention nodes from the estimated CPDAG. The local method of IDA can no longer be used for this purpose, since some combinations of locally valid parent sets of the intervention nodes may not yield a ``jointly valid" combination of parent sets. In \cite{NandyMaathuisRichardson14b} the authors proposed a semi-local algorithm for obtaining jointly valid parent sets from a CPDAG. The runtime of this semi-local algorithm is comparable to the runtime of the local algorithm in sparse settings. Moreover, the semi-local algorithm has the advantage that it (asymptotically) produces a multiset of joint intervention effects with correct multiplicities (up to a constant factor). It can therefore also be used in IDA if the multiplicity information is important.

JointIDA based on RRC or MCD was shown to be consistent in sparse high-dimensional settings.

\subsection{Application}\label{sec: validations IDA}

The IDA method is based on various assumptions, including multivariate Gaussianity, faithfulness, no hidden variables, and no feedback loops. In practice, some of these assumptions are typically violated. It is therefore very important to see how  the method performs on real data.

Validations were conducted in \cite{MaathuisColomboKalischBuehlmann10} on the yeast gene expression compendium of \cite{HughesEtAl00}, and in \cite{StekhovenEtAl12} on gene expression data of \emph{Arabidopsis Thaliana}. JointIDA was validated in \cite{NandyMaathuisRichardson14b} on the DREAM4 in silico network challenge \cite{MarbachEtAl09}. We refer to these papers for details. 

In the remainder, we want to highlight the severity of the order-dependence of the PC algorithm in high-dimensional settings (see Section \ref{sec: PC}), and also advocate the use of sub-sampling methods. We will discuss these issues in the context of the yeast gene expression data of \cite{HughesEtAl00}. These data contain both observational and experimental data, obtained under similar conditions. We focus here on the observational data, which contain gene expression levels of 5361 genes for 63
wild-type yeast organisms.

\begin{figure}[!ht]\centering%
  \subfigure[Black entries indicate edges occurring in the estimated skeletons using the PC algorithm, where each row in the figure corresponds to a different random variable ordering. The original ordering is shown as variable ordering 26. The edges along the x-axis are ordered from edges that occur in the estimated skeletons for all orderings, to edges that only occur in the skeleton for one of the orderings.  Red entries denote edges in the
  uniquely estimated skeleton using the PC-stable algorithm over the same 26
  variable orderings (shown as variable ordering 27).]{
     \includegraphics[width=0.45\textwidth,height=0.45\textwidth]{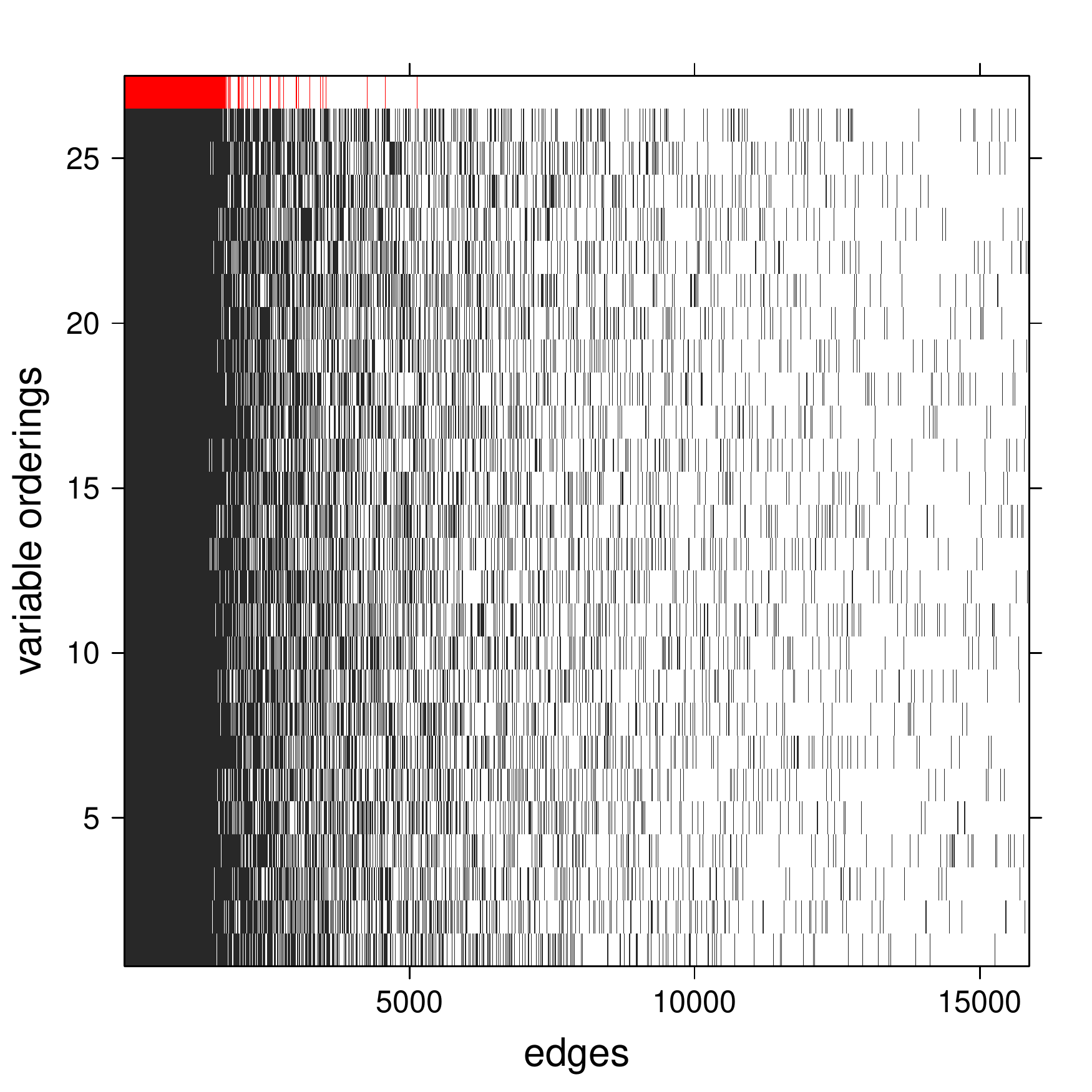}
     \label{fig.levelplot.2}
  }\qquad
   \subfigure[The step function shows the proportion of the 26 variable orderings
   in which the edges were present for the original PC algorithm, where the
   edges are ordered as in Figure \ref{fig.levelplot.2}. The red bars show
   the edges present in the estimated skeleton using the PC-stable algorithm.]{
     \includegraphics[width=0.45\textwidth,height=0.45\textwidth]{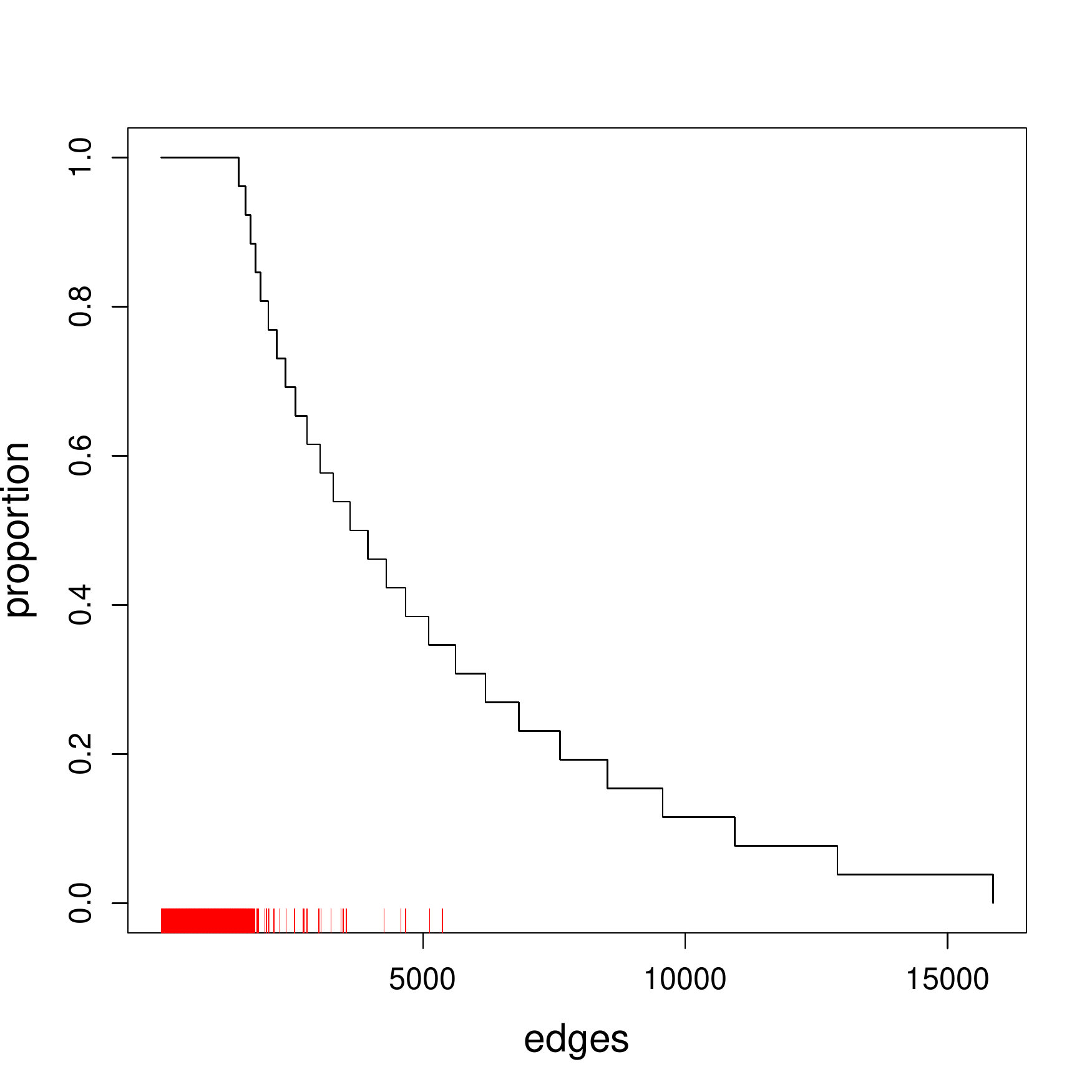}
     \label{fig.percplot.1}
  }
  \caption{Analysis of estimated skeletons of the CPDAGs for the yeast gene
    expression data \cite{HughesEtAl00}, using the PC and PC-stable
    algorithms with tuning parameter $\alpha=0.01$. The PC-stable algorithm yields an order-independent
    skeleton that roughly captures the edges that were stable among the
    different variable orderings for the original PC algorithm. Taken from \cite{ColomboMaathuis14}. }
  \label{fig.skeletanalysis.1}
\end{figure}

\begin{figure}[!ht]\centering%
    \includegraphics[scale=0.4,angle=0]{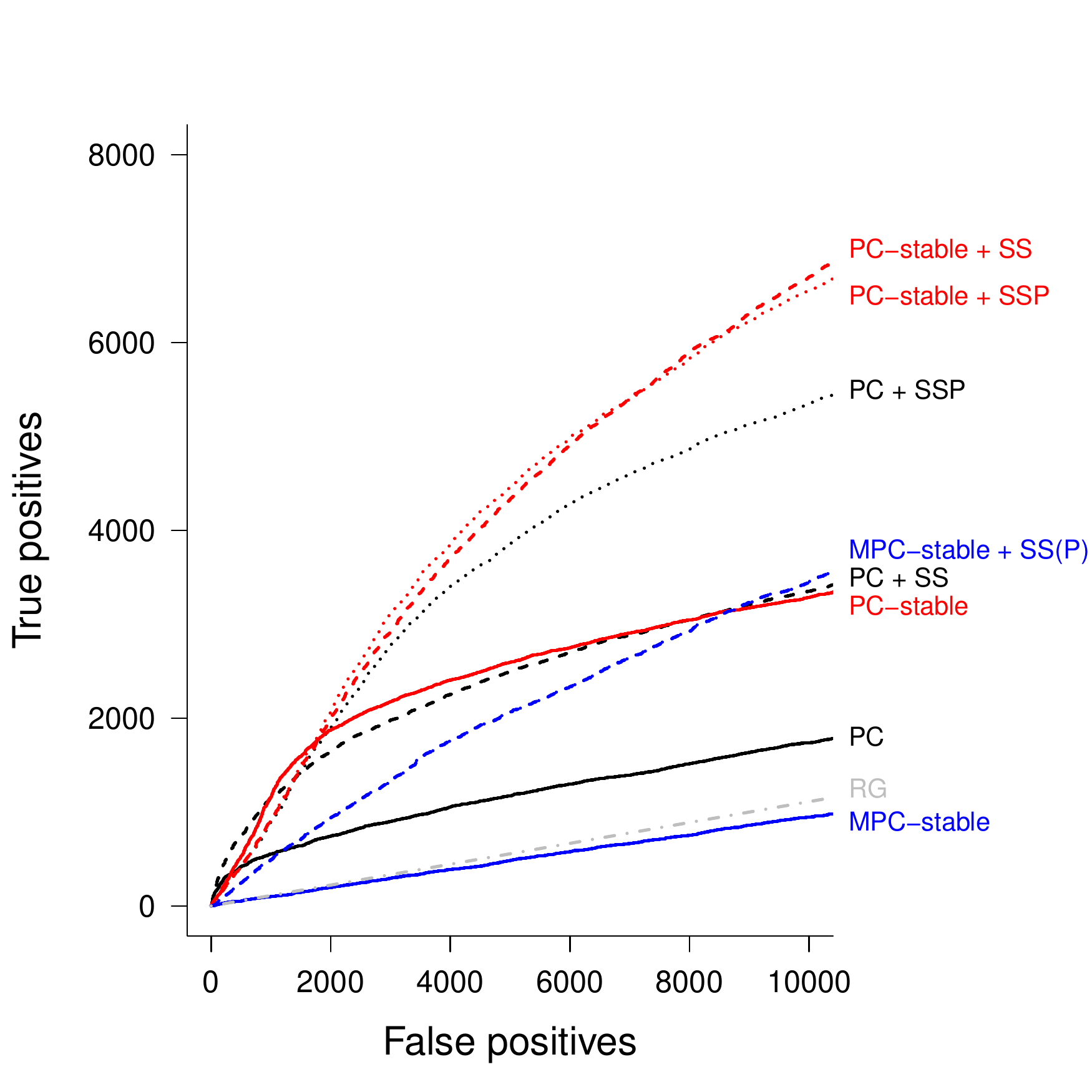}
    \caption{Analysis of the yeast gene expression data \cite{HughesEtAl00} with PC (black lines),
    PC-stable (red lines), and MPC-stable (blue lines), using the original ordering over
    the variables (solid lines), using 100 runs stability selection without
    permuting the variable orderings (dashed lines, labelled with ``+ SS"), and
    using 100 runs stability selection with permuting the variable
    orderings (dotted lines, labelled with ``+ SSP"). The grey line labelled as
    ``RG" represents random guessing. Taken from \cite{ColomboMaathuis14}.}
  \label{fig.hughesetal.stabsel}
\end{figure}

Let us first consider the order-dependence. The ordering of the columns in our $63 \times 5361$ observational data matrix should be irrelevant for our problem. But permuting the order of the columns (genes) dramatically changed the estimated skeleton. This is visualized in Figure \ref{fig.skeletanalysis.1} for 25 random orderings. Each estimated skeleton contained roughly 5000 edges. Only about 2000 of those were stable, in the sense that they occurred in almost all estimated skeletons. We see that PC-stable (in red) selected the more stable edges. Perhaps surprisingly, it did this via a small modification of the algorithm (and not by actually estimating skeletons for many different variable orderings). 

Next, we consider adding sub-sampling. Figure \ref{fig.hughesetal.stabsel} shows ROC curves for various versions of IDA. In particular, there are three versions of PC: PC, PC-stable and MPC-stable. Here PC-stable yields an order-independent skeleton, and MPC-stable also stabilizes the edge orientations. For each version of IDA, one can add stability selection (SS) or stability selection where the variable ordering is permuted in each sub-sample (SSP). We note that adding SSP yields an approximately order-independent algorithm. The best choice in this setting seems PC-stable + SSP.


\section{Extensions}\label{sec: extensions}

There are various extensions of the methods described in the previous sections. We only mention some directions here. \\

\noindent{\bf Local causal structure learning.} Recall from Section \ref{sec: determining post-intervention distribution} that we can determine the total effect of $X_i$ on $Y$ by adjusting for the direct causes, that is, by adjusting for the parents of $X_i$ in the causal graph. Hence, if one is interested in a specific intervention variable $X_i$, it is not necessary to learn the entire CPDAG. Instead, one can try to learn the local structure around $X_i$. Algorithms for this purpose include, e.g., \cite{TsamardinosEtAl03, Ramsey06, AliferisEtAl10a, AliferisEtAl10b}. \\

\noindent{\bf Causal structure learning in the presence of hidden variables and feedback loops.}
Maximal ancestral graphs (MAGs) can represent conditional independence information and causal relationships in DAGs that include unmeasured (hidden) variables \cite{RichardsonSpirtes02}. Partial ancestral graphs (PAGs) describe a Markov equivalence class of MAGs. PAGs can be learned from observational data. A prominent algorithm for this purpose is the FCI algorithm, an adaptation of the PC algorithm \cite{SpirtesEtAl93, SpirtesMeekRichardson95,SpirtesEtAl00,Zhang08-orientation-rules}. Adaptations of FCI that are computationally more efficient include RFCI and FCI+ \cite{ColomboEtAl12, ClaassenMooijHeskes13}. High-dimensional consistency of FCI and RFCI was shown by \cite{ColomboEtAl12}. The order-dependence issues studied in \cite{ColomboMaathuis14} (see Section \ref{sec: PC}) apply to all these algorithms, and order-independent versions can be easily derived. The algorithms FCI, RFCI and FCI+ are available in the R-package \texttt{pcalg} \cite{KalischEtAl12}. There is also an adaptation of LiNGAM that allows for hidden variables \cite{HoyerEtAl08}. Causal structure learning methods that allow for feedback loops can be found in \cite{Richardson96, MooijEtAl11, MooijHeskes13}.\\

\noindent{\bf Time series data.} Time series data are suitable for causal inference, since the time component contains important causal information. There are adaptations of the PC and FCI algorithms for time series data \cite{ChuGlymour08, EntnerHoyer10, Ebert-UphoffDeng12}. These are computationally intensive when considering several time lags, since they replicate variables for the different time lags.

Another approach for discrete time series data consists of modelling the system as a structural vector autoregressive (SVAR) model. One can then use a two-step approach, first estimating the vector autoregressive (VAR) model and its residuals, and then applying a causal structure learning method to the residuals to learn the contemporaneous causal structure. This approach is for example used in \cite{HyvarinenEtAl10}.

Finally, \cite{BrodersenEtAl15} proposed an approach based on Bayesian time series models, applicable to large scale systems. \\

\noindent{\bf Causal structure learning from heterogeneous data.}
There is interesting work on causal structure learning from heterogeneous data. For example, one can consider a mix of observational and various experimental data sets \cite{HauserBuehlmann12,PetersEtAl15}, or different data sets with overlapping sets of variables \cite{TillmanEtAl08,TriantafilouEtAl10}, or a combination of both \cite{TsamardinosEtAl12}. A related line of work is concerned with transportability of causal effects \cite{BareinboimPearl14}. \\

\noindent{\bf Covariate adjustment.}
Given a DAG and a set of intervention variables $\mathbf{X}$ and a set of target variables $\mathbf{Y}$, Pearl's backdoor criterion is a sufficient graphical criterion to determine whether a certain set of variables can be used for adjustment to compute the effect of $\mathbf{X}$ on $\mathbf{Y}$. This result was strengthened by \cite{ShpitserPearl06a} who provided necessary and sufficient conditions. In turn, this result was generalized by \cite{VanderZanderEtAl14} who provided necessary and sufficient conditions for adjustment given a MAG. Pearl's backdoor criterion was generalized to CPDAGs, MAGs and PAGs by \cite{MaathuisColombo14}. Finally, \cite{PerkovicEtAl15} provided necessary and sufficient conditions for adjustment in DAGs, MAGs, CPDAGs and PAGs.\\

\noindent{\bf Measures of uncertainty.}
The estimates of IDA come without a measure of uncertainty. (The regression estimates in IDA do produce standard errors, but these assume that the estimated CPDAG was correct. Hence, they underestimate the true uncertainty.)  Asymptotically valid confidence intervals could be obtained using sample splitting methods (cf. \cite{MeinshausenMeierBuehlmann09}), but their performance is not satisfactory for small samples. Another approach that provides a measure of uncertainty for the presence of direct effects is given by \cite{PetersEtAl15}. More work towards quantifying uncertainty would be highly desirable.

\section{Summary}\label{sec: discussion}

In this paper, we discussed the estimation of causal effects from observational data. This problem is relevant in many fields of science, since understanding cause-effect relationships is fundamental and randomized controlled experiments are not always possible. There is a lot of recent progress in this field. We have tried to give an overview of some of the theory behind selected methods, as well as some pointers to further literature.

Finally, we want to emphasize that the estimation of causal effects based on observational data cannot replace randomized controlled experiments. Ideally, such predictions from observational data are followed up by validation experiments. In this sense, such predictions could help in the design of experiments, by prioritizing experiments that are likely to show a large effect.

\bibliographystyle{plain}
\bibliography{../../MyBibliography}

\printindex

\end{document}